\documentstyle[10pt,aas2pp4]{article} 

\newcommand{\lalph}{\mbox{$L_m$--$\alpha$}} 
\newcommand{\lalphx}{\mbox{$L_m$--$\alpha$--$L_X$}} 
\newcommand{\etal}{{\em et al.\/}} 

\slugcomment{} 

\lefthead{Hudson \& Ebeling} 
\righthead{Brightest Cluster Galaxies} 

\begin{document} 
\title{The Environmental Dependence of Brightest Cluster Galaxies: 
Implications for Large-Scale Flows} 

\author{Michael J. Hudson\altaffilmark{1}} 
\affil{ 
Department of Physics \& Astronomy, University of Victoria, 
P.O. Box 3055, Victoria, B.C. V8W 3P6, Canada; 
hudson@uvastro.phys.uvic.ca 
} 
\altaffiltext{1}{CITA National Fellow} 

\and 

\author{Harald Ebeling\altaffilmark{2}} 
\altaffiltext{2}{{\em also:} Institute for Astronomy, 
2680 Woodlawn Drive, Honolulu, HI 96822, USA; ebeling@ifa.hawaii.edu} 
\affil{ 
Institute of Astronomy, Madingley Road, Cambridge CB3 0HA, 
UK 
} 

\begin{abstract} 
  In a much-noticed recent study Lauer and Postman (1994) found that
  the inertial frame defined by a sample of 119 nearby Abell clusters
  with $cz_{\sun} < 15000$ km~s$^{-1}$ showed a highly significant
  motion with respect to the cosmic microwave background (CMB) frame.
  We construct a subsample of their sample which comprises 64
  Abell/ACO clusters with X-ray luminosities from ROSAT and brightest
  cluster galaxy (BCG) photometry from Lauer and Postman. We find that
  both BCG metric luminosities and residuals from the \lalph\ relation
  of Lauer \& Postman are significantly correlated with the X-ray
  luminosity of the host cluster at the 99.6\% confidence level, in
  the sense that more X-ray luminous clusters have brighter BCGs.  The
  strength of this correlation increases with increasing X-ray
  luminosity and with increasing values of the structure parameter
  $\alpha$. Taking this correlation into account, we obtain a new
  distance indicator for BCGs, the \lalphx\ relation. Applying the
  \lalphx\ relation to our sample, we find that the frame defined by
  these clusters has a bulk motion of 494 km~s$^{-1}$ towards $l=
  285\arcdeg, b=47\arcdeg$ with respect to the CMB frame but the 95\%
  confidence range on the amplitude is 306 to 1419 km~s$^{-1}$. When
  the covariance of the components of the bulk motion is properly
  taken into account, these results are inconsistent with this frame
  being at rest in the CMB frame at the 98.6\% confidence level but
  are consistent with the 300--400 km~s$^{-1}$ amplitude flows found
  by other studies on scales $cz \lesssim 6000$ km~s$^{-1}$.  In order
  to obtain an estimate of the bulk flow on scales beyond local
  perturbations such as the ``Great Attractor'', we have also examined
  the subsample of 57 clusters with X-ray data and $cz_{\rm LG} >
  6000$ km~s$^{-1}$.  The random errors in the bulk motion are large
  due to the depth and small size of this sample.  We find that the
  bulk motion of the clusters in this shell with respect to the CMB
  frame is not statistically significant, but the 95\% confidence
  limits for the amplitude range from 27 to 2025 km s$^{-1}$.  The
  motion of this sample is also consistent with the motion found by
  Lauer and Postman. However, our analysis of all 107 Lauer and
  Postman BCGs with $cz_{\rm LG} > 6000$ km~s$^{-1}$ indicates that,
  even with no X-ray correction, the motion of these clusters with
  respect to the CMB frame is not significantly different from zero.
  Furthermore, the correction to the bulk motion of the subsample with
  the X-ray data goes in the sense of reducing the amplitude (by 663
  km~s$^{-1}$) and significance (from 98.8\% to 83.8\%) of its
  motion in the CMB frame, as well as reducing the internal
  inconsistency between its motion and that of the remainder of the
  Lauer and Postman sample with no X-ray data.  Claims of large-scale,
  large-amplitude bulk flows should therefore be regarded with caution
  until X-ray data become available for more clusters, or cluster
  distances are confirmed by independent methods.
\end{abstract} 


\section{Introduction} 
\label{sec:intro} 
% History of BCGs and LP result 
Brightest cluster galaxies (BCGs) have been used as cosmological 
probes since the pioneering work of Sandage and collaborators 
(\markcite{HMS}Humason, Mayall \& Sandage 1956; 
\markcite{San72a}\markcite{San72b}Sandage 1972a,b). Initially, BCGs 
were seen as a promising means of measuring $q_0$ (e.g. Sandage 
1972a,b). Recently, BCGs have been used to measure the large-scale 
streaming motion of the local Universe, with conflicting results 
(\markcite{San75}Sandage 1975; \markcite{JJC}James, Joseph \& Collins 
1987; \markcite{LC}Lucey \& Carter 1988). Lauer \& Postman (1994, 
hereafter \markcite{LP}LP; Postman \& Lauer 1995, hereafter 
\markcite{PL}PL), using a sample of BCGs in Abell/ACO clusters (Abell 
1958; Abell, Corwin \& Olowin 1989) with $cz_{\sun} < 15000$ km 
s$^{-1}$, found that the inertial frame defined by these clusters 
(hereafter ACIF) is moving with respect to the cosmic microwave 
background (CMB) frame at $689\pm178$ km~s$^{-1}$ which is significant 
at the greater than 99.99\% ($4\sigma$) level. Given that a 
large-scale flow of this amplitude is in conflict with most 
cosmological models at the 95-99\% confidence limit 
(\markcite{SCOLP}Strauss \etal\ 1995; \markcite{FW}Feldman \& Watkins 
1994; but see also \markcite{JK}Jaffe and Kaiser 1995), it is clearly 
important to re-examine this result and, in particular, the 
corrections applied to the BCG magnitudes as measured. 

% Corrections 
BCGs are not perfect standard candles and most workers have found that 
some form of correction to their magnitudes is necessary before they 
can be used as distance indicators. Sandage 
(\markcite{San72a}\markcite{San72b}1972a,b) applied both cluster 
richness and Bautz-Morgan (\markcite{BM}1970, BM) type corrections to 
BCG metric luminosities. Hoessel (\markcite{Hoe80}1980) found that 
taking into account the correlation between the luminosity $L_m$ 
within a metric aperture of radius $R_m$ and the logarithmic slope of 
the luminosity profile at that aperture, $\alpha \equiv 
(d\log(L_m)/d\log R)\vert_{R_m}$, led to reduced scatter and 
eliminated the need for cluster richness or BM corrections. However, 
Hoessel \& Schneider (\markcite{HS}1985) found that the correction 
``removes much of the richness and BM -- luminosity trends, but 
perhaps not all.'' Finally, \markcite{LP}LP used an \lalph\ relation 
similar to that of Hoessel (1980), again with no richness or BM 
corrections, in order to derive the bulk motion of the ACIF. 
\markcite{PL}PL showed that there is no correlation between Abell 
richness and residuals from their \lalph\ relation (although a visual 
examination of their Figure 7d suggests that the BCGs of Richness 
Class $\ge 2$ clusters may have slightly brighter residuals than 
poorer clusters). 

A correlation between the luminosity of the BCG and the environment of 
the host cluster may arise for different reasons. For instance, such a 
correlation may have an astrophysical basis in the sense that BCGs in 
richer environments grow more easily by cannibalising other cluster 
galaxies. Alternatively, such a correlation might arise from 
statistical considerations: if BCGs are simply the brightest members 
of a population drawn at random from a luminosity function, then 
clusters with more galaxies will tend to have brighter BCGs 
(\markcite{Sco57}Scott 1957; \markcite{Pee68}Peebles 1968). While this 
distinction is important for understanding the history of galaxy 
formation and mergers in clusters, it is immaterial for the main 
purpose of this paper -- we are concerned with {\em any empirical}\/ 
correlation and its consequences for large-scale motions. 

However, the richness or projected galaxy density may not be the most 
sensitive probe of the cluster environment. In particular, projection 
effects can cause the two-dimensional richness to be a poor estimator 
of the true three-dimensional galaxy density around the BCG 
(\markcite{Luc83}Lucey 1983; \markcite{Sut88}Sutherland 1988; 
\markcite{Haa96}van Haarlem 1996). Only the latter, however, can be 
expected to be correlated with the BCG luminosity. Selection of 
clusters by the X-ray emission from the gaseous intra-cluster medium 
has several advantages over optical selection. First, the existence of 
diffuse gas at temperatures of typically $10^{7-8}$ K guarantees that 
the cluster is a bound system with a deep potential well. Second, the 
X-ray emission is unlikely to be contaminated by foreground/background 
groups which are projected onto the cluster. This is because the X-ray 
volume emissivity is proportional to the square of the gas density and 
thus much more peaked than the projected galaxy distribution by which 
clusters have traditionally been selected and defined at optical 
wavelengths. Finally, if clusters are selected from different optical 
catalogues in different hemispheres (e.g. Abell vs. ACO catalogues), 
then a systematic difference in cluster richness (e.g.\ 
\markcite{SZV}Scaramella \etal\ 1991) would bias the bulk motion if 
the properties of the BCG are indeed correlated with richness. The use 
of X-ray data from an all-sky survey eliminates all these biases and 
allows more physical parameters such as the X-ray gas temperature or 
the X-ray luminosity to be used to parametrise a cluster's 
richness. Indeed, \markcite{Edg91}Edge (1991) found a strong 
correlation between the cluster X-ray temperature and BCG magnitudes, 
and a slightly weaker correlation between X-ray luminosity and BCG 
magnitude. 

%Outline 
The goals of this work are twofold: first, to determine if BCG 
photometric properties are correlated with the X-ray luminosities of 
their host clusters; and second, to investigate the impact of such 
correlations on the derived large-scale flow field. In Section 
\ref{sec:data}, we introduce the X-ray sample, and in Section 
\ref{sec:xcorr} we show that a strong correlation of this type does 
indeed exist. In Section \ref{sec:motion}, we investigate the 
implications of this correlation for large-scale flows by performing a 
simultaneous fit to the parameters of the \lalphx\ relation and the 
motion of the Local Group (LG). Throughout we adopt $H_0 = 80$ km 
s$^{-1}$ Mpc$^{-1}$ and $q_0 = 0.5$ when calculating luminosities, but 
quote distances in units of km~s$^{-1}$. 

\section{Data} 
\label{sec:data} 
% X-ray data and corrections 
The sample of the X-ray brightest Abell-type clusters (XBACs) of 
Ebeling \etal\ (\markcite{Ebe96}1996) is an X-ray flux-limited ($f_X > 
5.0 \times 10^{-12}$ erg cm$^{-2}$ s$^{-1}$ in the 0.1 -- 2.4 keV 
band) sample of all Abell/ACO (Richness $\ge 0$) clusters detected in 
the ROSAT All-Sky Survey (RASS). 56 of the 119 clusters in the ACIF 
sample are also contained in the XBACs sample. A further 17 of the 
clusters in the ACIF sample have been detected in the RASS but fall 
below the flux limit of the published XBACs sample (Ebeling, private 
communication). The RASS X-ray luminosities of these additional 17 
clusters remain proprietary data of MPE and are not available for this 
study. A search of the ROSAT data archive, however, uncovered another 
12 detections of ACIF clusters in deeper pointings with the PSPC (the 
same detector that was used during the RASS) thus bringing our total 
sample of X-ray detected ACIF clusters to 68. Five of these 68 are 
classified as double clusters in the X-ray. As a comparison between 
cluster X-ray fluxes from the RASS and fluxes for the same clusters 
obtained from pointed PSPC observations showed the two data sets to be 
in excellent agreement (Ebeling \etal\ 1996), we are confident that 
the merging of cluster detections from the RASS and pointed 
observations does not introduce any systematic bias. If a cluster from 
the XBACs sample has also been observed in a PSPC pointing, we adopt 
the flux from the pointed observation as the latter goes deeper in all 
cases thus yielding better photon statistics. (We have confirmed that 
our results are unaffected if we use only the RASS data.) All X-ray 
fluxes are accurate to typically 10 to 20 per cent and have been 
corrected for foreground Galactic absorption. Note that the fluxes are 
dominated by emission from the intracluster medium and not by 
individual sources such as, for instance, the BCG or contaminating 
AGN. In the case of the pointed observations this was ensured by 
explicitly removing all flux from point sources from the overall 
emission; for the RASS detections where the photon statistics are 
usually too poor to allow such an individual treatment a statistical 
correction was applied to the total cluster emission (see Ebeling 
\etal\ 1996 for details). 

% Optical data: from PL calculating alpha and gamma 
On the optical side, we use the BCG photometry of \markcite{PL}PL 
(their Table 3). We fit a quadratic form in log(aperture) to the 
tabulated photometry yielding the parameters $M_L$, $\alpha_L$ 
(evaluated at the 10 $h^{-1}$ kpc radius aperture assuming the cluster 
is at rest in the LG frame) and $\alpha' \equiv (d\alpha/d\log 
R)\vert_{R_m}$. The derivative $\alpha$ and second derivative 
$\alpha'$ allow us to determine, by Taylor series expansion, $L_m$ and 
$\alpha$ for any assumed BCG distance and corresponding $10 h^{-1}$ 
kpc metric aperture. Extinction and $k$ corrections are as in PL. Of 
the 68 clusters that our X-ray sample has in common with the ACIF 
list, we reject the following clusters for which the positions of the 
BCG and the X-ray centroid were found to be discrepant: A189, a 
multiple system (\markcite{ZGH}Zabludoff \etal\ 1993) in which the BCG 
of LP corresponds to the group at $cz=9925$ km~s$^{-1}$ while the 
X-ray centroid is coincident with the foreground system centred on NGC 
533 at $cz = 5544$ km~s$^{-1}$; A1228, also a multiple system 
(\markcite{ZGH}Zabludoff \etal\ 1993) in which the BCG of LP is in the 
``A'' group at $cz = 10674$ km~s$^{-1}$, and the X-ray centroid is 
centred on UGC 06394 in the ``B'' group at $cz = 12715$ km~s$^{-1}$; 
A3560 for which the LP BCG is NGC 5193 at $cz = 3644$ km~s$^{-1}$ and 
the main cluster is in the background at $cz = 14840$ km~s$^{-1}$ 
(\markcite{VCS}Vettolani \etal\ 1990); and A3869 in which the LP BCG 
is NGC 7249 at $cz = 12005$ km~s$^{-1}$ and the X-ray centroid is 
coincident with the cluster APM 222041.3-552848 at $z=0.078$ 
(\markcite{DEM}Dalton \etal\ 1994) as also noted by Ebeling \etal\ 
(1996). In the five cases where two X-ray subclusters are associated 
with a given Abell/ACO cluster, we use again the component with the 
best positional agreement with the BCG (A548a, A1631a, A2197b, A2572a 
and A3528b) in order to ensure that the optical and X-ray parameters 
used in our study are indeed associated with the same physical system. 
We refer to the frame defined by this sample as the XACIF 
(X-ray--Abell cluster inertial frame). Table \ref{tab:data} lists the 
optical and X-ray properties of the XACIF sample. 

{\bf Table 1 follows the References.}

In our analysis of the bulk motion below, we shall also consider the 
statistically independent subsample of clusters for which X-ray data 
are unavailable, because they have X-ray fluxes below the flux limit 
of the XBACs sample and have not been observed in PSPC pointings 
either. We refer to this as the NOX sample. Note that many of the 
NOX clusters are detected in the RASS, but their fluxes have not been 
released by MPE. 

\begin{figure*}
\figurenum{1}
\epsscale{2.0}
\plotone{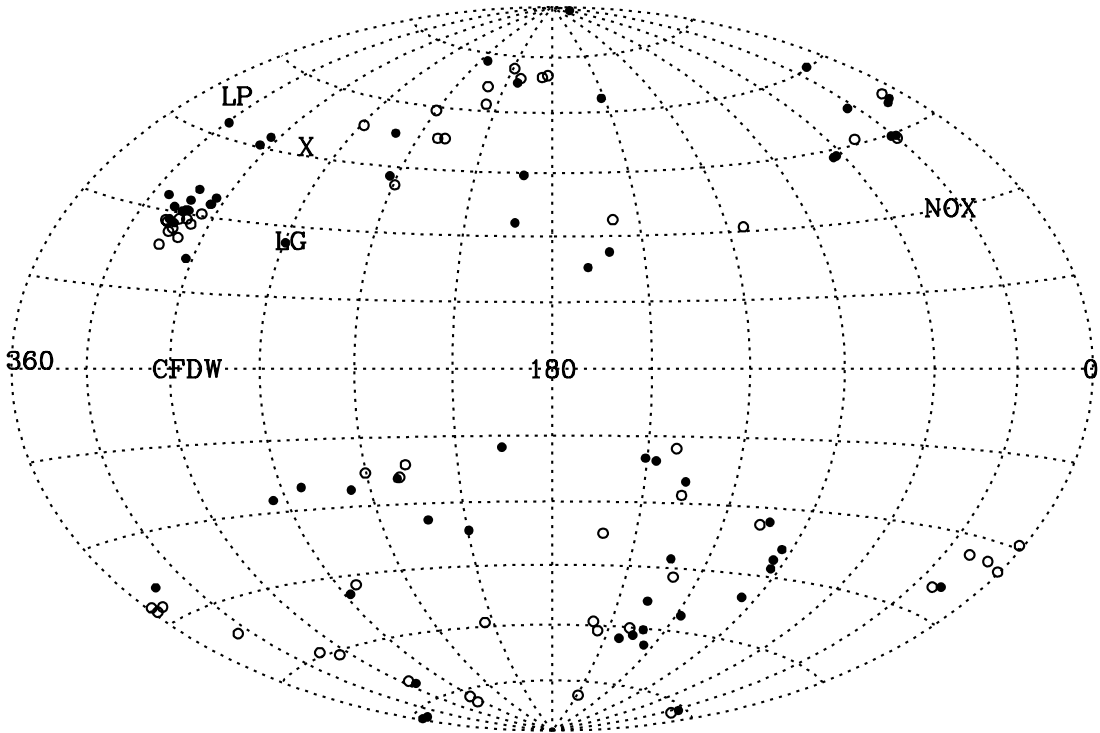}
\caption{% 
The distribution of clusters on the sky in an Aitoff projection of 
Galactic coordinates. Clusters with and without X-ray data are 
indicated by the filled and open circles respectively. The labels 
give the directions of the bulk motion in the CMB frame for samples 
of LP (indicated by ``LP'') and Courteau \etal\ 1993 (``CFDW''), for 
the XACIF (``X'') and NOX samples (``NOX'') discussed in Section 
\protect{\ref{sec:motion}}, and for the Local Group (``LG''). 
\label{fig:sky} 
} 
\end{figure*}

Figure \ref{fig:sky} shows the distribution of clusters on the sky. 
The clusters with X-ray and without X-ray data are shown by the filled 
and open circles respectively. 

\section{The Environmental Dependence of BCGs} 
\label{sec:xcorr} 

\begin{figure*}
\figurenum{2}
\plotone{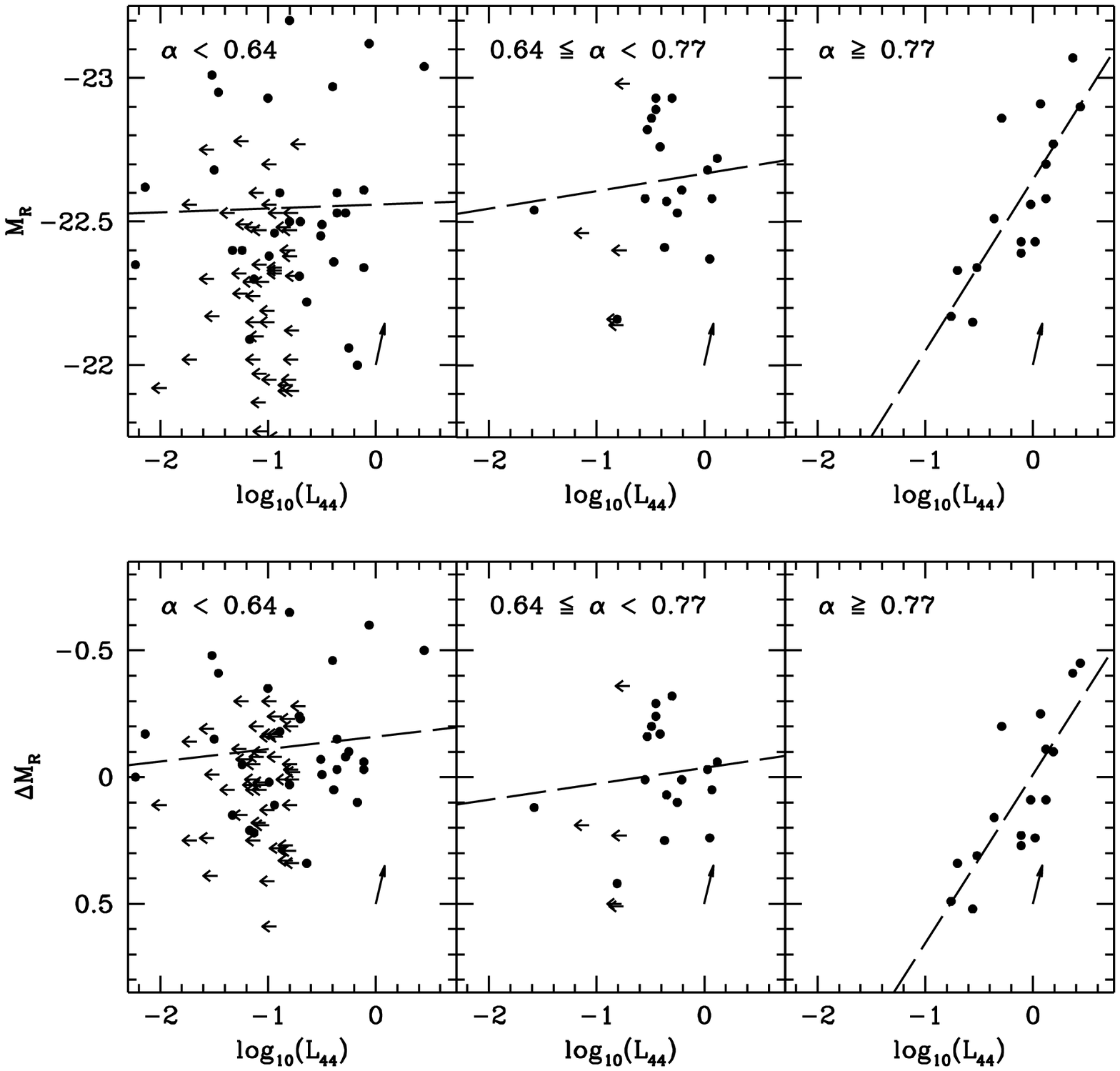}
\caption{% 
The correlations between the BCG properties $M$ and $\alpha$ and the 
host cluster's X-ray luminosity (LG frame). The top panels show BCG 
metric magnitude as a function of $\log(L_{44})$ whereas the lower 
panels show the residuals from the \lalph\ relation, $\Delta 
M(\alpha)$, as a function of $\log(L_{44})$. The data in the XACIF 
sample are indicated by solid circles. BCGs for which we have only 
upper limits on the cluster X-ray luminosity are indicated by 
horizontal arrows. The sample is subdivided by $\alpha$: the left 
panels show the half of the sample with $\alpha < 0.66$, the middle 
panels show the second quartile with $0.64 \le \alpha < 0.77$ and 
the right panels show the top quartile ($\alpha \ge 0.77$). The 
dashed lines show the best fit to the BCGs with X-ray detections in 
each subsample. The correlation found for high values of $\alpha$ is 
highly significant. The arrow on the lower right of each panel 
indicates the change in parameters if the distance of a cluster is 
increased by 10\% (twice the typical error due to peculiar 
velocities). 
\label{fig:xraymag} 
} 
\end{figure*}

Figure \ref{fig:xraymag} shows the correlations between the BCG 
properties $M$ and $\alpha$ and the host cluster's X-ray luminosity. 
The top panels of Figure \ref{fig:xraymag} show BCG magnitude as a 
function of $L_X$, whereas the lower panels show the residuals from 
the \lalph\ relation, $\Delta M(\alpha)$, as a function of $L_X$. The 
sample is subdivided by $\alpha$: the left panels show the half of the 
sample with $\alpha < 0.64$, the middle panels show the quartile with 
$0.64 \le \alpha < 0.77$ and the right panels show the top quartile 
($\alpha \ge 0.77$). The data in the XACIF sample are indicated by 
solid circles. BCGs for which we have only upper limits on the cluster 
X-ray luminosity are indicated by horizontal arrows. The dashed lines 
show the best fit to each subsample. Note that the X-ray coverage of 
the ACIF sample is essentially complete for high X-ray luminosity/high 
$\alpha$ systems. It is clear that for large $\alpha$ both $M$ and 
$\Delta M(\alpha)$ are significantly correlated with $L_X$. 
Linear and rank correlation statistics applied to the detections 
indicate that this correlation is significant at the $\gtrsim 95$\% 
level for $\alpha \gtrsim 0.6$. The significance of these 
correlations remains the same whether we assume that the ACIF is at 
rest with respect to the LG or to the CMB frame. It should also be 
noted that this correlation is not the result of peculiar velocities 
perturbing both $L_X$ and $\Delta M(\alpha)$ since a typical peculiar 
velocity of 500 km~s$^{-1}$ at a distance of 10000 km~s$^{-1}$ changes 
$\log(L_X)$ by only 0.042 and $\Delta M$ by $\sim 0.075$ mag. 

In order to assess whether these correlations remain significant when 
we include upper limits, we have calculated the generalized Kendall's 
$\tau$ statistic (\markcite{IFN}Isobe, Feigelson and Nelson 1986) 
using the ASURV package, Rev 1.2 (available from code@stat.psu.edu). 
This statistic tests for the existence of a rank correlation allowing 
for one or both variables to be limits or detections. When we cut the 
sample, we find significant ($\ge 95$\%) correlations for all 
subsamples for which $\alpha_{\rm min} \ge 0.6$. In fact, the 
correlation for the complementary $\alpha < 0.6$ subsample is 
marginally significant (at the 93.3\% confidence level). 

\begin{figure*}
\figurenum{3}
\plotone{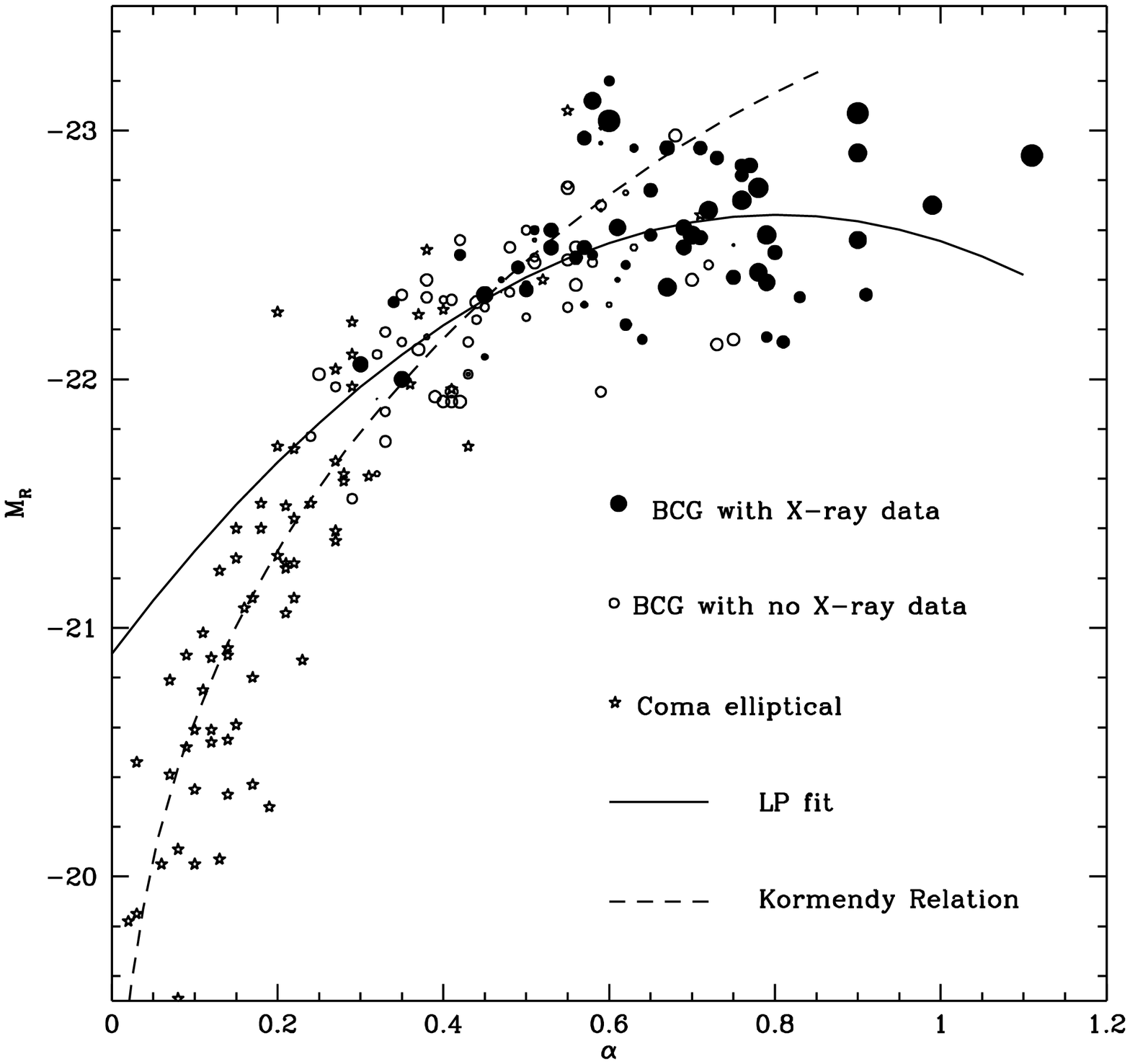}
\caption{% 
The \lalph\ relations for BCGs and giant ellipticals. Filled circles 
are BCGs in clusters with X-ray data and open circles are BCGs with no 
X-ray data. The radius of the circle is proportional to $\log L_{42}$ 
(the measured X-ray luminosity or the upper limit, for filled and open 
circles respectively). Ellipticals in Coma are indicated by stars. 
These have been mapped onto the $M_{R}$--$\alpha$ diagram by assuming 
that they have $R^{1/4}$ law profiles. The solid line is the \lalph\ 
relation of LP. The dashed curve is the projection of the Kormendy 
relation followed by giant elliptical galaxies. 
\label{fig:fpcoma} 
} 
\end{figure*}

Similar to PL's Figure 1 our Figure \ref{fig:fpcoma} shows the \lalph\ 
diagram of the ACIF sample with the clusters with and without X-ray 
information plotted as filled and open circles respectively. In order 
to compare the properties of BCGs with those of cluster giant 
ellipticals, we have mapped the ellipticals in Coma 
(\markcite{LGCT}Lucey \etal\ 1991) onto the \lalph\ diagram assuming 
that they follow an $R^{1/4}$ law and adopting a mean colour of $V-R_c 
= 0.56$. Also plotted is the \markcite{Kor77}Kormendy (1977) relation 
(a projection of the Fundamental Plane) of \markcite{GLB}Guzman et 
al.\ (1993) for the same galaxies. Note the good continuity between 
the distribution of giant ellipticals and the lower $\alpha$ half of 
the LP BCG sample. The Kormendy relation appears to be an acceptable 
fit to both giant ellipticals and BCGs up to $\alpha \sim 0.55$, 
whereas for higher values of $\alpha$ the \lalph\ relation deviates 
from the Kormendy relation. Also, it appears from Figure 
\ref{fig:xraymag} that, for $\alpha \lesssim 0.6$, the \lalph\ 
residuals do not correlate strongly with the X-ray luminosity of the 
host cluster, whereas above this value we find an increasingly strong 
dependence on $L_X$. Finally it is worth noting that there also exists 
a highly significant correlation between $\alpha$ and $L_X$ in the 
sense that larger $\alpha$ galaxies are found in more X-ray luminous 
clusters. A simple interpretation of all the above observations is 
that low-$\alpha$ BCGs are typically found in low $L_X$ (and hence 
typically poor) clusters and follow the same Fundamental Plane 
relations as lower-ranked cluster ellipticals independent of the 
cluster environment. High-$\alpha$ BCGs, on the other hand, are 
typically found in rich clusters and have photometric properties which 
depend on the cluster's X-ray luminosity and presumably on its mass. 

To summarize, we have shown that, for large $\alpha$ ($\gtrsim 0.6$) 
BCGs, there is a significant correlation between both the BCG 
magnitudes themselves and the residuals of the \lalph\ relation and 
$L_X$. For small $\alpha$ ($\lesssim 0.6$) BCGs, the correlation 
between residuals of the \lalph\ relation and $L_X$ is only marginally 
significant (at the 93.3\% confidence level) and, in any case, weaker 
than that found for the large-$\alpha$ BCGs. The X-ray correction to 
an individual BCG aperture magnitude can be quite large (e.g.\ $\sim 
0.5$ mag for the X-ray luminous cluster A3571). A difference between 
the mean X-ray luminosities of clusters on opposite sides of the sky 
could thus translate into a spurious bulk flow if the X-ray correction 
is neglected. 

\section{The Bulk Motion} 
\label{sec:motion} 

\subsection{Method} 
\label{sec:method} 
In order to investigate the effects of the X-ray correlation on the
derived bulk flow, it is necessary to determine a \lalphx\ relation
and re-derive the motion of the sample. We fit simultaneously both the
parameters of the \lalphx\ relation, denoted by ${\bf c}$, and the
motion of the LG with respect to the sample, ${\bf L}$. Our method of
fit is to minimize the aperture magnitude residuals
(\markcite{Col95}Colless 1995)
\begin{equation} 
\chi^2 = \sum_i^{N} \left(\frac{\Delta M(r_i({\bf L}),{\bf c})} 
{\sigma_m(r_i({\bf L}))}\right)^2 
\label{eq:chi}
\end{equation} 
simultaneously as a function of $\bf{c}$ and $\bf{L}$. Note that the
motion of the LG with respect to the XACIF sample fixes the distance
(in units of km~s$^{-1}$), $r_i$, of each BCG via the relation $r_i =
cz_{{\rm LG},i} + {\bf L} \cdot\hat{\bf r}_i$, where $z_{{\rm LG},i}$
is the observed redshift of the $i^{\rm th}$ cluster in the LG frame.

The residuals from this fit are given by
\begin{equation} 
\Delta M(r,{\bf c}) = M(r) - \overline{M(r,{\bf c})} 
\label{eq:res} 
\end{equation} 
where the predicted magnitude is 
\begin{eqnarray} 
\overline{M(r,{\bf c})} & = & c_0 + c_1\,\alpha + c_2\,\alpha^2 + 
 c_3 \, \log(L_{44}) 
\nonumber \\
& &  + c_4 \, \alpha \log(L_{44}) + 
c_5 \, \alpha^2 \log(L_{44}) 
\,. 
\end{eqnarray} 
$L_{44}$ is the cluster X-ray luminosity in units of $10^{44}$ ergs
s$^{-1}$ in the 0.1 -- 2.4 keV band.  Note that $M$, $\alpha$ and
$\log(L_{44})$ are all non-linear functions of $r$ and hence of
$\bf{L}$.

For the NOX sample (for which X-ray data are unavailable) we follow an
identical procedure to that described above, except that we use an
\lalph\ relation of the same functional form as LP, i.e.\ we set
$c_3$, $c_4$ and $c_5$ equal to zero. For completeness we also
rederive the motion of the original ACIF sample neglecting all X-ray
data and using the \lalph\ relation.

The total scatter in magnitudes is $\sigma^2_m = \sigma_0^2 +
\left(\frac{d\Delta M}{dr} \, \sigma_v\right)^2$, where $\sigma_0$ is
the intrinsic scatter in magnitudes about the \lalphx\ relation, and
$\sigma_v$ is the dispersion in peculiar velocity around the
best-fitting flow model. This scatter in velocity arises from two
sources. Firstly, we have peculiar velocities due to structures within
the sampled volume that are not accounted for because we model the
flow as a simple bulk motion. Gramann \etal\ (\markcite{GBCG}1995)
estimate the one-dimensional velocity dispersion of clusters to be
approximately 300 km~s$^{-1}$. Secondly, there will be a contribution
from observational errors in the cluster redshifts. \markcite{LP}LP
estimate the errors in measured redshift to be approximately 184 km
s$^{-1}$. Adding these in quadrature, we adopt $\sigma_v = 350$ km
s$^{-1}$. The velocity scatter contributes to the $\sigma_m$ of nearby
clusters but is negligible for the more distant clusters for which the
error is dominated by $\sigma_0$. Our final results are not very
sensitive to the choice of $\sigma_v$.

Using equation \ref{eq:res}, it is straightforward to calculate the
change in magnitude residuals for a given change in log distance.
\begin{equation} 
\frac{d\Delta M}{d\log r} = 
\frac{dM(r)}{d\log r} - 
\frac{d\overline{M(r,{\bf c})}}{d\log r} 
\label{eq:ddeldr} 
\end{equation} 
with 
\begin{equation} 
\frac{dM(r)}{d\log r} = 
-2.5\,[2-\alpha] 
\label{eq:dmdr} 
\end{equation} 
and 
\begin{eqnarray} 
\frac{d\overline{M(r,{\bf c})}}{d\log r} 
& = &  -\alpha'\,\left[\,c_1 + 2 \, c_2 \, \alpha + c_4 \, \log(L_{44}) 
\right. \nonumber \\ 
& & \hspace{0.75cm} \left. 
+ 2 c_5\, \alpha\, \log(L_{44})\,\right] 
\nonumber  \\ 
& & + 2\,\left[\,c_3 + c_4 \, \alpha + c_5\, \alpha^2\,\right] 
\label{eq:dmpreddr}
\end{eqnarray} 
where we have used $d\alpha/d\log r = -\alpha'$ and $d\log
L_{44}/d\log r = 2$.\footnote{ While the calculation of the magnitude
  residual (equation \ref{eq:res}) is always performed
  relativistically, the expressions for the derivatives (equations
  \ref{eq:dmdr} \& \ref{eq:dmpreddr}) neglect relativistic terms which
  arise from (a) the change in luminosity with redshift and (b) the
  change in the angular diameter of the $10 h^{-1}$ metric aperture.
  For a typical BCG the change in $M(r)$ due to the relativistic terms
  over the typical distance error is 0.008 mag (to be compared with
  the \lalph\ scatter of 0.244 mag).  The effect of including the
  relativistic terms would be to reduce the estimated errors, but due
  to the low mean redshift ($z \sim 0.025$) of the sample, the effect
  on the fractional distance error is small: a factor $\sim 0.97$.

  As the distance of a galaxy changes, its position in the \lalph\ 
  diagram changes too.  We approximate its true path, which is mildly
  parabolic, by a straight line.  This is valid if the distance errors
  are small.  For the typical galaxy the change in magnitude residual
  due to this second-order term is very small: 0.002 mag, two orders
  of magnitude smaller than the scatter 0.244 mag.  Furthermore, since
  this correction can have either sign depending on whether the
  distance is over- or underestimated, it will tend to cancel out.  We
  are therefore justified in neglecting this second-order term.
}

The fractional distance error is then given by
\begin{equation} 
\sigma_r = \ln(10) 
\left(\frac{d\Delta M}{d\log r}\right)^{-1} \sigma_0 
\, 
\label{eq:fracerr} \\ 
\end{equation} 
and the peculiar velocity error is the quadrature sum of the distance
error and $\sigma_v$. Note that $\alpha$ is a distance-dependent
quantity. For an $R^{1/4}$ profile, $\alpha$ is monotonically related
to $\log R_e$ (\markcite{GLCP}Graham \etal\ 1996). When calculating
their peculiar velocity {\em errors\/} \markcite{LP}LP considered only
equation (\ref{eq:dmdr}) corresponding to the change in the {\em
  observed\/} aperture magnitude with distance. However, because
$\alpha$ is a (weakly) distance-dependent variable it is important not
to neglect the $d\alpha/d\log r = -\alpha'$ terms which correspond to
the change in the {\em predicted\/} magnitude as a function of
distance (i.e.\ equation \ref{eq:dmpreddr}).

For nearly all of the BCGs in the ACIF sample, the result of including
this term is that the distance error increases. This is particularly
important at low $\alpha$ where the slope of the \lalph\ relation is
steep, and hence the change in predicted magnitude for a change in
$\alpha$ is large. For example, consider a small $\alpha = 0.3$ galaxy
with a typical value of $\alpha' = -0.5$. Using the parameters of LP's
\lalph\ relation, LP's equation (2) gives a distance error of 13\%
whereas equation (\ref{eq:fracerr}) gives an error of 23\% (neglecting
the X-ray dependent terms). Consequently, in our analysis of the ACIF
sample the mean distance error is $18.7\%$ with an rms of $6\%$,
roughly independent of $\alpha$, whereas LP's $\alpha$-dependent
distance error has a mean of 15.8\%\footnote{ For the mean redshift
  $\sim 0.025$, the effect of including relativistic terms discussed
  in the previous footnote is to reduce our \lalph\ fractional error
  from 18.7\% to 18.2\%.  When one outlying BCG (A3374) with large
  distance error is excluded, the mean fractional distance error is
  17.7\%, in good agreement with the scatter of 17.3\% obtained by
  comparing estimated redshifts with observed redshifts on a
  galaxy-by-galaxy basis (PL).}.  The net result is that while our
ACIF ${\bf L}$ agrees with LP's value, the {\em errors\/} on the three
components are systematically slightly larger than those quoted by LP.
We have confirmed that, if we neglect the distance dependence of
$\alpha$ by setting $\alpha'$ to zero, we obtain the smaller errors of
LP.

The X-ray luminosity is also a distance-dependent quantity, and we
find that the \lalphx\ relation steepens for large $L_X$. (Note that
in Figure \ref{fig:xraymag}, the line of best fit steepens and becomes
closer in slope to the arrow indicating the effect of distance
errors.) Consequently, very large $\alpha$ BCGs in the most X-ray
luminous clusters are poor distance indicators.  The distance error
for the \lalphx\ relation is typically in the range from 11\% to 35\%
with a median of 17\%.

For the $\chi^2$ minimization (equation \ref{eq:chi}), the effective
weight of a BCG in the bulk flow solution is proportional to the
inverse square of its peculiar velocity error, which is the fractional
error (equation \ref{eq:fracerr}) times the BCG distance.  Thus, for
our solutions, high and low $\alpha$ galaxies contribute with
aproximately equal weights, and the weight drops with distance
approximately as $r^{-2}$. The exceptions are the very large $\alpha$
BCGs in the most X-ray luminous clusters, which, by virtue of their
large fractional errors, have low weight.

\begin{figure*}
\figurenum{4}
\plotone{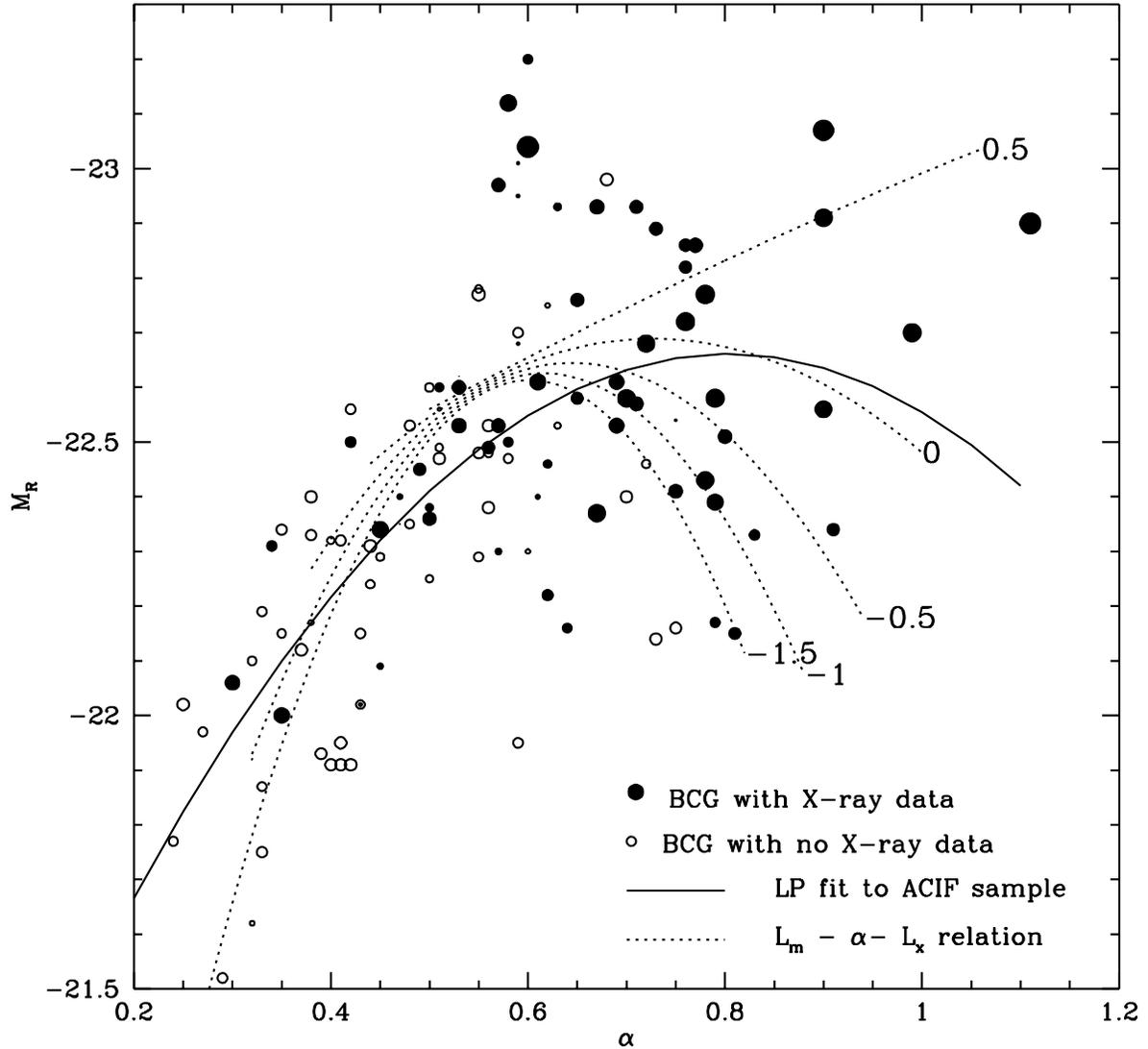} 
\caption{% 
The \lalph\ and \lalphx\ relations for BCGs. Filled circles are BCGs 
in clusters with X-ray data and open circles are BCGs with no X-ray 
data. The radius of the circle is proportional to $\log L_{42}$ (the 
measured X-ray luminosity or the upper limit for filled and open 
circles respectively). The solid line is the \lalph\ relation of 
LP. The dotted curves show the \lalphx\ relation for values of 
$\log(L_{44})$ ranging from --1.5 to 0.5. 
\label{fig:lalphx}
} 
\end{figure*}

In order to determine $\sigma_0$, we fix $\sigma_v$ and adjust
$\sigma_0$ so that (for the best-fitting solution) we obtain $\chi^2$
equal to the number of degrees of freedom (the number of clusters less
the nine free parameters). This yields $\sigma_0 = 0.231$, which is
less than the value 0.253 obtained when we fit the \lalph\ relation to
the same 64 clusters. (The $\sigma_m$ of 0.244 found by
\markcite{LP}LP for the \lalph\ relation alone was for all 119
clusters. However, LP noted that the scatter increases for the $\alpha
< 0.6$ subsample where most of the clusters in the XACIF sample are
found.) We adopt $\sigma_0 = 0.231$ in order to evaluate the errors on
the parameters. For the best-fit solution, the parameters are $c_0
=-21.219$, $c_1 =-4.046$, $c_2 = 2.783$, $c_3 =-1.605$, $c_4 = 5.724$,
$c_5 = -5.139$. Figure \ref{fig:lalphx} compares the \lalphx\ relation
to the \lalph\ relation of LP.

% Comparison to quality of fit without X-ray data 
If we set $c_3$, $c_4$ and $c_5$ to zero, thereby ignoring the X-ray
data, we find that $\chi^2$ increases by 13.5 for an increase of only
3 in the number of degrees of freedom. This check confirms that the
correlation with X-ray luminosity is highly significant (99.6\%
confidence). Indeed, it is worth noting that the reduction in $\chi^2$
due to the X-ray correction, parametrized by the $c_3$, $c_4$ and
$c_5$, is far more significant than that due to the choice of bulk
flow, parametrized by the three components of $\bf L$, as will be
shown below.

\subsection{Monte Carlo Experiments} 
\label{sec:monte} 

To assess whether the geometry of the XACIF sample biases the flow 
solution, we have performed Monte Carlo (MC) experiments with mock 
data for the XACIF data sample. For each such mock data set, we use 
the measured position, redshift and profile shape (which determines 
$\alpha$ and its derivative for any distance) of each BCG in the XACIF 
sample, assume a bulk flow and parameters of the \lalphx\ relation, 
and then randomly generate X-ray luminosities using the 
$\alpha$--$L_X$ relation and aperture magnitudes using the \lalphx\ 
relation. We find that the scatter in the bulk flow parameters from 
one MC realization to another is in excellent agreement with the 
errors inferred from our $\chi^2$ minimization. We find only a very 
small level of ``geometry bias'' in our recovered flow solutions: the 
results differ from the input values by typically 10--20 km~s$^{-1}$ 
in each component. This is a factor $\sim 20$ smaller than the random 
errors. 

We have also examined whether our results are affected by a 
Malmquist-like bias due to the X-ray flux limit. Since the ACIF is 
believed to be volume-limited, we can use it as the underlying cluster 
density field from which X-ray flux-limited samples are drawn. We 
assign X-ray luminosities and aperture magnitudes as described above 
to the entire ACIF sample and then impose a flux limit which matches 
that of the XBACs sample in order to generate mock XACIF samples. The 
resulting bias, which includes both Malmquist and geometry biases, is 
at the 10--20 km~s$^{-1}$ level. We conclude that for our analysis 
both geometry and Malmquist biases are small and so can be neglected. 

\subsection{Bulk Flow Results} 
\label{sec:results} 

\begin{deluxetable}{llrrrrrrrrr} 
\tablenum{2}
\scriptsize
\tablecaption{Bulk Flow Solutions \label{tab:flow}} 
\tablecolumns{11} 
\tablehead{ 
\colhead{Sample} & 
\colhead{Fit$^{a}$} & 
\colhead{$\sigma_0$} & 
\colhead{$N$} & 
\colhead{$r_{\rm eff}^{b}$} & 
\colhead{$L_x^{c}$} & 
\colhead{$L_y^{c}$} & 
\colhead{$L_z^{c}$} & 
\colhead{$|{\bf F}|^{d}$} & 
\colhead{$l^{d}$} & 
\colhead{$b^{d}$}\\ 
\colhead{}& 
\colhead{}& 
\colhead{}& 
\colhead{}& 
\colhead{km~s$^{-1}$} & 
\colhead{km~s$^{-1}$} & 
\colhead{km~s$^{-1}$} & 
\colhead{km~s$^{-1}$} & 
\colhead{km~s$^{-1}$} & 
\colhead{} & 
\colhead{} 
} 
\startdata 
\cutinhead{Full redshift range} 
{\bf XACIF } & {\bf XF } & {\bf 0.231 } & {\bf 64 } & {\bf 7356 } & {\bf 
-175 $\pm$ 462 } & {\bf 22 $\pm$ 390 } & {\bf -361 $\pm$ 365 } & {\bf 
494$^{+ 925}_{- 188}$ } & {\bf 285 } & {\bf 47} \\ 
XACIF & NXA & 0.243 & 64 & 7684 & 
$ -207 \pm 458$ & $ 118 \pm 403$ & $ -378 \pm 372$ & 
$ 618^{+ 888}_{- 234}$ & 286 & 43 \\ 
NOX & NXF & 0.203 & 55 & 8327 & 
$ -984 \pm 368$ & $-1227 \pm 490$ & $ -414 \pm 322$ & 
$ 1183^{+ 1220}_{- 844}$ & 35 & 30 \\ 
ACIF & NXF & 0.243 & 119 & 8031 & 
$ -521 \pm 303$ & $ -353 \pm 339$ & $ -303 \pm 250$ & 
$ 593^{+ 673}_{- 282}$ & 339 & 47 \\ 
\cutinhead{$cz_{\rm LG} > 6000$ km~s$^{-1}$} 
{\bf XACIF } & {\bf XF } & {\bf 0.238 } & {\bf 57 } & {\bf 10373 } & {\bf 
144 $\pm$ 566 } & {\bf 254 $\pm$ 575 } & {\bf -356 $\pm$ 399} & {\bf 
492$^{+ 1533}_{- 465}$ } & {\bf 258 } & {\bf 37} \\ 
XACIF & NXA & 0.243 & 57 & 10507 & 
$ 63 \pm 549$ & $ 682 \pm 543$ & $ -489 \pm 389$ & 
$ 1155^{+ 1251}_{- 675}$ & 266 & 31 \\ 
NOX & NXF & 0.205 & 50 & 10209 & 
$-1025 \pm 471$ & $-1363 \pm 566$ & $ -270 \pm 341$ & 
$ 1153^{+ 1365}_{- 850}$ & 39 & 22 \\ 
ACIF & NXF & 0.246 & 107 & 10434 & 
$ -428 \pm 392$ & $ -267 \pm 436$ & $ -246 \pm 271$ & 
$ 308^{+ 1037}_{- 221}$ & 326 & 46 
\enddata 
\tablenotetext{a}{Fits are coded as follows: XF --- \lalphx\ relation, 
parameters free; NXF --- \lalph\ relation parameters are free; 
NXA --- \lalph\ relation, parameters are fixed to those of the 
ACIF solution } 
\tablenotetext{b}{The effective depth of the sample.} 
\tablenotetext{c}{The motion of the Local Group with respect to the 
cluster sample.} 
\tablenotetext{d}{Motion of sample in the CMB frame. The amplitude 
quoted has been ``error-bias corrected''. Errors represent the 95\% 
range in the raw amplitude with the direction fixed. Note that 
these errors should not be used to determine whether the flow is 
compatible with a given model. This can only be accomplished with 
the full covariance matrix.} 
\end{deluxetable} 

Table \ref{tab:flow} gives the solutions for the motion of the LG, 
$\bf{L}$, in Galactic Cartesian coordinates with respect to the XACIF, 
NOX and ACIF samples. The directions of the bulk flow motions found 
for the respective samples are marked in Figure \ref{fig:sky}. 

Converting the motion of the LG with respect to the sample to the
motion of the sample with respect to the CMB, we find that the XACIF
clusters have a motion of 867 km~s$^{-1}$ towards $l=285\arcdeg$,
$b=47\arcdeg$ in the CMB frame. However, as noted by \markcite{LP}LP,
the amplitude of the flow is biased upwards by the random errors.
After correction for ``error biasing'' by subtracting the errors from
the observed amplitude in quadrature, the best estimate of the
amplitude of the bulk flow is in the CMB frame is 494 km~s$^{-1}$, but
the 95\% confidence range on the amplitude is 306 to 1419 km~s$^{-1}$.
In contrast, the error-bias corrected bulk motion of the NOX sample is
1183 km~s$^{-1}$ towards $l=35\arcdeg, b=30\arcdeg$, but again the
range in amplitude is large: 339 to 2403 km~s$^{-1}$. We caution that
the one-dimensional errors quoted above are given only to indicate the
approximate level of the random errors. They should {\em not\/} be
used to compare the motions of the different subsamples or to assess
the significance of the motion with respect to the CMB frame. The
proper way to perform such comparisons is to use the full covariance
matrices of the errors, as described in Section \ref{sec:compare}
below.

Clusters within 6000 km~s$^{-1}$ carry a very large weight in the
fits, and are predominantly located in two superclusters
(Hydra-Centaurus and Perseus-Pisces), where peculiar velocities may be
particularly high. Furthermore, this volume is known to have a mean
bulk motion of $360 \pm 40$ km~s$^{-1}$ (Courteau \etal\
\markcite{CFDW}1993, hereafter CFDW). In order to obtain an
independent estimate of the bulk motion on large scales, we also
analyze samples with $cz_{\rm LG} > 6000$ km~s$^{-1}$ which we refer
to as the outer shell. For the XACIF sample in the outer shell, we
obtain a motion of 492 km~s$^{-1}$ towards $l=258\arcdeg$,
$b=37\arcdeg$, but now the 95\% range in amplitude is very large: 27
to 2025 km~s$^{-1}$. If, for the same sample, the \lalph\ relation
obtained from the full ACIF sample is used as a distance indicator, we
obtain a motion of 1155 km~s$^{-1}$ towards $l=266\arcdeg$,
$b=31\arcdeg$, with a 95\% lower limit on the amplitude of 480 km
s$^{-1}$. The vector corresponding to the X-ray correction has an
amplitude of 455 km~s$^{-1}$ in the direction $l=281\arcdeg$,
$b=17\arcdeg$ (close to the direction of the flow in the CMB
frame). Clearly, for the outer shell, the X-ray correction has made a
large difference to the amplitude and significance of the inferred
flow. The NOX sample in the outer shell also appears to show evidence
of motion: the 95\% lower limit on its amplitude is 303 km~s$^{-1}$
towards $l=39\arcdeg$, $b=22\arcdeg$ ($110\arcdeg$ from the XACIF
motion). The difference vector between the XACIF and NOX outer shell
bulk motions has an amplitude of $1998\pm810$ km~s$^{-1}$. Note that
the XACIF and NOX motions disagree strongly in their X and Y
components which differ by 1169 km s$^{-1}$ and 1617 km~s$^{-1}$
respectively. On the other hand, the Z components, which disagree with
CFDW, agree well with each other.  If, instead, we use the \lalph\
relation obtained from the full ACIF sample, the disagreement between
the bulk motions of the XACIF and NOX subsamples in the outer shell is
even worse: the bulk motions differ by $2306\pm846$ km~s$^{-1}$.

\subsection{The Statistical Significance and Consistency 
  of the Flow Solutions}
\label{sec:compare} 

\subsubsection{Independent Samples}
In order to assess whether our results are consistent with independent 
samples and with various assumed flow models (e.g.\ one in which the 
sample is at rest in the CMB frame), it is necessary to use the full 
covariance matrix of the bulk motion errors. To compare two 
independent samples, with motions ${\bf L}_1$ and ${\bf L}_2$, and 
corresponding covariance matrices ${\bf C}_1$ and ${\bf C}_2$, we 
calculate 
\begin{equation} 
\chi^2 = ({\bf L}_1 - {\bf L}_2)^{\rm T} \left({\bf C}_1 + {\bf 
C}_2\right)^{-1} ({\bf L}_1 - {\bf L}_2) 
\end{equation} 
and compare the result to a $\chi^2$ distribution with 3 degrees of 
freedom. Note that when comparing two peculiar velocity samples with 
different sky coverage and effective depth, we do not expect the 
measured bulk flows, ${\bf L}_1$ and ${\bf L}_2$, to be identical even 
in the limit of no measurement errors due to the different window 
functions (cf.\ \markcite{WF}Watkins \& Feldman 1995). Therefore, the 
confidence with which we conclude that two samples are inconsistent 
will in general be slightly overestimated. However, in the specific 
case of the comparison between the XACIF and NOX samples discussed 
below, it is worth noting that the volumes sampled are very similar in 
depth and sky coverage (see Table \ref{tab:flow} and Figure 
\ref{fig:sky}), therefore we expect these samples to partake of the 
same bulk flow. If we wish to test the hypothesis that, for example, 
a given sample is at rest with respect to the CMB frame, then ${\bf 
C}_2$ is set to zero and $\bf{L}_2$ is set to the motion of the LG 
with respect to the CMB frame. Table \ref{tab:compare} compares the 
consistency of the motion of the XACIF, NOX and ACIF samples with the 
CMB frame and with the motion found by CFDW. 

\begin{deluxetable}{llrrr} 
\tablenum{3}
\tablecaption{Statistical Comparison of Flow Solutions \label{tab:compare}} 
\tablewidth{0pt} 
\tablecolumns{5} 
\tablehead{ 
\colhead{} & 
\colhead{Fit$^{a}$} & 
\colhead{${P_{\rm CMB}}^b$} & 
\colhead{${P_{\rm CFDW}}^c$} & 
\colhead{${P_{\rm NOX}}^d$} 
} 
\startdata 
\cutinhead{Full redshift range} 
XACIF & XF & 0.014 & 0.142 & 0.185 \\ 
XACIF & NXA & 0.005 & 0.078 & 0.231 \\ 
NOX & NXF & 0.032 & 0.050 \\ 
ACIF & NXF & 0.004 & 0.063 \\ 
\cutinhead{$cz_{\rm LG} > 6000$ km~s$^{-1}$} 
XACIF & XF & 0.162 & 0.312 & 0.098 \\ 
XACIF & NXA & 0.012 & 0.053 & 0.046 \\ 
NOX & NXF & 0.079 & 0.062 \\ 
ACIF & NXF & 0.114 & 0.247 
\enddata 
\tablecomments{Probability that two given flow solutions are 
consistent.} 
\tablenotetext{a}{Fits are coded as in Table \ref{tab:flow}} 
\tablenotetext{b}{The probability that the sample is a rest with respect 
to the CMB frame.} 
\tablenotetext{c}{The probability that the sample has the motion found by 
Courteau \etal\ (1993).} 
\tablenotetext{d}{The probability that X-ray selected sample agrees with the solution of the corresponding NOX sample.} 
\end{deluxetable} 

The first result of this analysis is that we reject the hypothesis 
that the XACIF sample is at rest with respect to the CMB frame at the 
98.6\% ($2.5\sigma$) confidence level. However, this is not surprising 
given the fact that the motion of this sample is dominated by clusters 
at a typical distance of only 7300 km~s$^{-1}$, and the fact that the 
local volume to 60 $h^{-1}$ Mpc is known to have a $\sim 350$ km 
s$^{-1}$ bulk motion (CFDW). In fact the XACIF motion is consistent 
with the bulk motion found by CFDW. 

On the largest scales, for the outer shell ($cz_{\rm LG} > 6000$ km 
s$^{-1}$), there is no significant (83.8\%) evidence of bulk motion 
with respect to the CMB frame, although for this shell the random 
errors are large (400 -- 600 km~s$^{-1}$ per vector component). 
Despite these large errors, note that had we neglected the X-ray 
correction and used instead the \lalph\ relation derived from the 
whole ACIF sample as a distance indicator, we would have concluded 
that the XACIF clusters in the outer shell sample had a highly 
significant (98.8\% confidence) motion with respect to the CMB frame. 

In order to further investigate the effect of the X-ray correction on 
the derived bulk flows, we compare the statistically independent XACIF 
and NOX samples. (It is not valid to compare, for example, the motion 
of XACIF sample with that of either the full ACIF sample or that the 
XACIF clusters using the \lalph\ relation because these samples are 
not independent, but see below for further discussion of this issue.) 
We remind the reader that a significant correlation between BCG 
magnitude and $L_X$ was found for the (on average more X-ray luminous) 
clusters in the XACIF sample, whereas a much weaker dependence on the 
X-ray luminosity of the host cluster is expected of the optical 
properties of the BCGs in the (on average much less X-ray luminous) 
clusters in the NOX sample (cf.\ Figure~\ref{fig:xraymag}). Under the 
null hypothesis that the X-ray correction is unimportant, we thus 
expect the XACIF and NOX samples to obey the same \lalph\ relation and 
to yield independent and mutually consistent estimates of the bulk 
flow. However, when we compare the bulk motions of the XACIF and NOX 
samples using the \lalph\ relation derived from the whole sample, we 
find that in the outer shell the bulk motions are mutually 
inconsistent at the 95.4\% confidence level. The X, Y and Z components 
of the bulk motions differ by 884, 2111 and 284 km~s$^{-1}$ 
respectively. Given that LP found no significant disagreement when 
they cut the sample by $\alpha$ or redshift, the XACIF--NOX 
disagreement highlights the importance of X-ray selection and the 
resulting correction to the derived bulk flow. 

We now consider solutions for the XACIF motion using the \lalphx\ 
relation, which allows us to correct for the X-ray dependence of the
sample. For the NOX sample, we allow the parameters of the \lalph\ 
relation to be free. This does not remove all of the X-ray dependence
of this sample, but allows us to make a correction to the mean X-ray
luminosity of the sample (which must be lower than that of the XACIF
sample given that most of the XACIF clusters where drawn from an X-ray
flux limited sample). As noted above, for the outer shell, the XACIF
motion is now consistent with being at rest in the CMB frame. The
discrepancy between the XACIF and NOX bulk motions is still present
but its significance is reduced to the 90.2\% confidence level. We
suggest that this remaining small discrepancy is caused by the lack of
X-ray correction in the solution for the NOX sample which, owing to
its low content in high-$\alpha$ BCGs, is, however, less strongly
affected than the XACIF sample.

\subsubsection{Correlated Samples}

We would also like to assess the statistical significance of the
difference between the bulk flow of the XACIF sample with and without
the X-ray correction applied, and the difference between the motion of
the XACIF sample and that of the ACIF sample. These comparisons cannot
be made using the methods described above because these samples are
not independent. Indeed, for the XACIF sample, the residuals from the
\lalph\ and \lalphx\ relations are highly correlated. The simplest way
to properly account for this correlation is to use the Monte Carlo
(MC) simulations described above.  In these simulations, we assume a
bulk motion and \lalphx\ relation, and generate MC realizations of the
cluster X-ray luminosities and BCG magnitudes. We can then calculate
bulk flows for each MC realization both with and without the X-ray
data included in the fit.  This allows us to determine the covariance
matrix of the vector difference between two bulk flow solutions in a
way which accounts for the correlated residuals.

We find that the observed vector difference between the ACIF and 
X-ray corrected XACIF motions in the outer shell is consistent with 
the MC covariance matrix derived from the differences between the 
bulk motion of mock ACIF realizations (using the \lalph\ relation) 
and the bulk motion of flux-limited X-ray subsamples (using the 
\lalphx\ relation). Therefore, we conclude that, in the outer 
shell, the X-ray corrected bulk motion of the XACIF sample is 
consistent with that derived from the ACIF sample 
without any X-ray correction. 

When we compare the bulk flow vectors of MC XACIF realizations with
and without the X-ray term, we find that vector differences of
amplitude $\sim 500$ km~s$^{-1}$ are not uncommon. These vector
differences, however, are randomly oriented with respect to the bulk
motion vector in the CMB frame, as expected since, in the MC
simulations, the cluster X-ray luminosities are uncorrelated with the
clusters' positions on the sky. For the actual XACIF sample, however,
the vector difference is nearly opposite in direction to the sample's
motion in the CMB frame, with the result that the amplitude of the
flow, which is 1155 km~s$^{-1}$ before the X-ray correction is
applied, is reduced to 492 km~s$^{-1}$ when the X-ray data are
included.  Such a large reduction in amplitude occurs in $\sim$ 25\%
of all cases in the MC simulations but such good alignment occurs in
only 2.5\% of all cases.  We interpret this as evidence that the high
amplitude motion ($1153_{-675}^{+1251}$ km~s$^{-1}$ where the errors
give the 95\% range) of the XACIF sample obtained with the \lalph\ 
relation alone (i.e. without the X-ray correction) is systematically
biased.

\subsection{Discussion} 
\label{sec:implications} 
We have shown that the X-ray correction to the BCG \lalph\ relation is
statistically significant and that the motion of the XACIF sample in
the outer shell ($cz_{\rm LG}> 6000$ km s$^{-1}$) is consistent both
with no motion with respect to the CMB and with the motion of the ACIF
sample found by LP.  The implications of this correlation for
large-scale flows can only be properly addressed when X-ray data
become available for the whole ACIF sample. Given the fact that, while
these data exist, they are presently unavailable for almost half of
the clusters in the ACIF sample, how do we interpret our results?

The conservative approach is to assume that, given the relevance of
the X-ray correction, one can safely use as distance indicators only
those BCGs that have measured cluster X-ray luminosities, and hence
that, on large scales, the flow solutions are consistent with the most
conservative interpretation that these systems are at rest in the CMB
frame.

On the other hand, an unfortunate consequence of the small size of the
XACIF sample is that the errors on the bulk motion are large so that
we also cannot reject the (albeit less conservative) hypothesis that
there is a $\sim 600$ km~s$^{-1}$ bulk flow as found by LP. The
following facts, however, argue against the existence of
large-amplitude flows beyond 6000 km~s$^{-1}$.  Firstly, the \lalphx\
distance indicator reduces the amplitude of the XACIF flow by 663
km~s$^{-1}$ and brings the XACIF sample into better agreement with the
CMB frame compared to the \lalph\ relation applied to the same BCGs.
A reduction of this amplitude is unlikely to occur if the X-ray
correction vector is randomly oriented with respect to the motion, as
would be required under the null hypothesis that there is no
systematic variation of the average cluster X-ray luminosity across
the sky.  Secondly, the motions of the XACIF and NOX samples in the
outer shell differ by $2306\pm846$ km~s$^{-1}$ and are inconsistent
with each other at the 95\% confidence level when no X-ray correction
is applied to either subsample.  Finally, even when we consider the
outer shell ACIF sample with no X-ray correction, we find that this
sample is still marginally consistent with being at rest in the CMB
frame (we would reject this at only the 89\% confidence level). Thus,
independent of the unknown X-ray correction to the NOX subsample, the
evidence for flows beyond 6000 km~s$^{-1}$ is weak.

Until all of these issues can be clarified by the addition of further
X-ray data, we prefer the conservative conclusion that there is no
evidence for large-amplitude large-scale motion beyond 6000
km~s$^{-1}$.

% Future prospects 
Given the large random errors on the XACIF bulk motion, it would 
clearly be valuable to obtain X-ray data for a larger sample of the 
ACIF clusters. However, caution should be exercised before applying 
the \lalphx\ relation to still deeper cluster samples. Typical 
distance errors increase from $\approx 17\%$ at the median $L_X$ in 
the XACIF sample to $\approx 30\%$ at highest X-ray luminosities due 
to the steepening of the \lalphx\ relation and the fact that $L_X$ is 
distance-dependent. Consequently, extending the \lalphx\ relation in 
order to measure the bulk flow in larger volumes is expected to be 
doubly difficult. The results of Edge (\markcite{Edg91}1991) indicate 
that BCG metric luminosity is better correlated with X-ray temperature 
than with X-ray luminosity. Furthermore, the fractional distance error 
does not increase with $T_X$ (whereas it does with $L_X$) because the 
X-ray temperature is distance-independent. A more promising approach 
would thus be to collect X-ray temperatures for a large all-sky sample 
of nearby clusters. 

\section{Conclusions} 
\label{sec:conc} 

Using the subset of \markcite{LP}LP clusters which have X-ray
luminosities from the XBACs sample of Ebeling \etal\ 
(\markcite{Ebe96}1996) or from pointed ROSAT PSPC observations, we
have shown that, for large-$\alpha$ BCGs, both the metric luminosities
and residuals from the \lalph\ relation are significantly correlated
with the X-ray luminosity of the host cluster at the 99.6\% confidence
level.

We have included $L_X$ as an additional parameter in the BCG relation
and re-derived the bulk motion of the XACIF sample. Although we cannot
rule out the possibility that the sample is at rest with respect to
the CMB at better than the 98\% confidence level, the motion of the
XACIF sample is fully consistent with the lower amplitude (300 -- 400
km~s$^{-1}$) flows found by other workers (CFDW) on scales of 6000 km
s$^{-1}$.

On larger scales ($cz_{\rm LG} > 6000$ km~s$^{-1}$), the random errors
in the derived bulk flow are large. The motion of XACIF sample is
consistent with the most conservative hypothesis that these clusters
are at rest in the CMB frame but also with the large amplitude motion
found by LP. However, even when the X-ray data are excluded from the
fit, the 107 clusters in the Lauer and Postman ACIF sample in the
outer shell do not have a significant bulk motion with respect to the
CMB.  Furthermore, we note that use of the \lalphx\ relation
introduces a systematic correction to the bulk flow of the XACIF
sample which is comparable to the random errors and goes in the sense
of reducing both the amplitude and the significance of its motion in
the CMB frame.  Had we used the \lalph\ relation employed by LP, we
would have incorrectly found that this sample had a highly significant
(98.8\%) motion with respect to the CMB. We would also have found that
the bulk motions of the XACIF sample and the remainder of the LP
sample disagreed by $2306\pm846$ km~s$^{-1}$ which is significant at
the 95\% confidence level.

We conclude that the evidence for flows beyond 6000 km~s$^{-1}$ is
weak and that claims of large-scale, large-amplitude flows should be
regarded with caution until further X-ray data become available.

\acknowledgements 

We thank the authors of Ebeling \etal\ (1996), and there in particular 
Dr.\ H.\ B\"ohringer, for making their X-ray cluster sample available 
to us prior to publication. Thanks are also due to Alastair Edge for 
many helpful discussions and to F.D.A. Hartwick for comments on an 
earlier version of the manuscript. We are indebted to the referee, 
Marc Postman, whose criticism helped to improve this paper. HE 
acknowledges financial support from an EARA fellowship and SAO 
contract SV4-64008.

\clearpage
\begin{deluxetable}{lrrrrrr} 
\tablenum{1}
\tablewidth{0pc} 
\tablecaption{The X-ray--Abell cluster sample \label{tab:data}} 
\tablehead{ 
\colhead{Name} & 
\colhead{$M_L$} & 
\colhead{$\alpha_L$} & 
\colhead{$\alpha'$} & 
\colhead{$\log(L_{44})$} & 
\colhead{$\Delta M_L$} & 
\colhead{$\sigma_r$} 
} 
\startdata 
A0076 & --22.493 & 0.55 & --0.09 & --0.50 & 0.110 & 0.151 \nl 
A0119 & --22.719 & 0.76 & --0.21 & 0.12 & --0.004 & 0.200 \nl 
A0147 & --22.493 & 0.42 & --0.38 & --0.69 & --0.155 & 0.204 \nl 
A0168 & --22.535 & 0.57 & --0.47 & --0.36 & 0.087 & 0.169 \nl 
A0193 & --22.615 & 0.68 & 0.30 & --0.21 & 0.058 & 0.169 \nl 
A0194 & --22.401 & 0.60 & --0.72 & --1.33 & 0.239 & 0.160 \nl 
A0195 & --22.398 & 0.46 & --0.47 & --1.24 & 0.014 & 0.229 \nl 
A0262 & --22.144 & 0.81 & 0.09 & --0.56 & 0.342 & 0.250 \nl 
A0376 & --22.525 & 0.69 & --0.30 & --0.25 & 0.144 & 0.170 \nl 
A0407 & --22.314 & 0.84 & --0.98 & --0.70 & 0.056 & 0.111 \nl 
A0496 & --22.576 & 0.78 & --0.51 & 0.12 & 0.141 & 0.209 \nl 
A0533 & --22.383 & 0.50 & --0.34 & --0.99 & 0.139 & 0.177 \nl 
A0539 & --22.362 & 0.50 & --0.17 & --0.39 & 0.172 & 0.155 \nl 
A0548a & --22.453 & 0.49 & --0.26 & --0.51 & 0.061 & 0.164 \nl 
A0569 & --22.354 & 0.47 & --0.81 & --2.23 & 0.041 & 0.850 \nl 
A0576 & --22.058 & 0.29 & --0.77 & --0.25 & --0.032 & 0.606 \nl 
A0671 & --22.934 & 0.71 & --0.04 & --0.45 & --0.293 & 0.176 \nl 
A0779 & --22.946 & 0.59 & --0.21 & --1.46 & --0.309 & 0.155 \nl 
A0957 & --22.821 & 0.76 & --0.17 & --0.53 & --0.242 & 0.183 \nl 
A1060 & --22.176 & 0.80 & --0.68 & --0.76 & 0.271 & 0.132 \nl 
A1139 & --22.296 & 0.58 & --0.30 & --1.12 & 0.335 & 0.159 \nl 
A1185 & --22.458 & 0.61 & --1.19 & --0.94 & 0.186 & 0.167 \nl 
A1314 & --22.497 & 0.58 & 0.09 & --0.80 & 0.134 & 0.147 \nl 
A1367 & --22.529 & 0.53 & --0.85 & --0.28 & 0.050 & 0.206 \nl 
A1631a & --22.587 & 0.65 & --0.60 & --0.55 & 0.068 & 0.160 \nl 
A1644 & --22.704 & 0.98 & --0.64 & 0.12 & --0.083 & 0.425 \nl 
A1656 & --23.041 & 0.60 & --0.47 & 0.45 & --0.392 & 0.174 \nl 
A1736 & --23.118 & 0.58 & --0.64 & --0.06 & --0.487 & 0.177 \nl 
A1836 & --22.679 & 0.58 & --0.51 & --1.50 & --0.047 & 0.164 \nl 
A1983 & --22.311 & 0.34 & --0.64 & --0.71 & --0.256 & 0.529 \nl 
A2052 & --22.557 & 0.89 & --0.55 & --0.02 & 0.046 & 0.244 \nl 
A2063 & --22.393 & 0.79 & --0.85 & --0.11 & 0.258 & 0.176 \nl 
A2107 & --22.863 & 0.77 & --0.51 & --0.28 & --0.241 & 0.173 \nl 
A2147 & --22.370 & 0.66 & --0.72 & 0.05 & 0.314 & 0.182 \nl 
A2151 & --22.412 & 0.75 & --0.26 & --0.37 & 0.210 & 0.179 \nl 
A2197b & --23.011 & 0.59 & --0.17 & --1.51 & --0.375 & 0.154 \nl 
A2199 & --22.769 & 0.78 & --0.60 & 0.19 & --0.035 & 0.216 \nl 
A2572a & --22.605 & 0.53 & 0.30 & --0.36 & --0.027 & 0.131 \nl 
A2589 & --22.430 & 0.78 & --0.17 & --0.11 & 0.228 & 0.202 \nl 
A2593 & --22.509 & 0.80 & --1.11 & --0.36 & 0.061 & 0.137 \nl 
A2634 & --22.765 & 0.65 & --0.26 & --0.41 & --0.104 & 0.162 \nl 
A2657 & --21.998 & 0.34 & --0.60 & --0.17 & 0.208 & 0.270 \nl 
A2717 & --22.342 & 0.91 & 0.21 & --0.52 & --0.088 & 0.694 \nl 
A2877 & --23.199 & 0.60 & --0.30 & --0.80 & --0.556 & 0.159 \nl 
A3144 & --22.096 & 0.44 & --0.30 & --1.17 & 0.258 & 0.194 \nl 
A3376 & --22.615 & 0.61 & --0.34 & --0.11 & 0.038 & 0.164 \nl 
A3389 & --22.604 & 0.51 & --0.47 & --0.90 & --0.061 & 0.187 \nl 
A3395 & --22.429 & 0.78 & --0.60 & 0.02 & 0.262 & 0.197 \nl 
A3526 & --22.883 & 0.72 & --0.64 & --0.45 & --0.248 & 0.156 \nl 
A3528b & --22.934 & 0.67 & --0.04 & --0.30 & --0.267 & 0.166 \nl 
A3530 & --22.861 & 0.76 & 0.09 & --0.49 & --0.272 & 0.207 \nl 
A3532 & --22.677 & 0.72 & --0.09 & 0.03 & 0.021 & 0.182 \nl 
A3556 & --22.966 & 0.57 & --0.43 & --0.40 & --0.344 & 0.167 \nl 
A3558 & --23.067 & 0.89 & --0.60 & 0.37 & --0.235 & 0.383 \nl 
A3559 & --22.926 & 0.62 & 0.00 & --1.00 & --0.280 & 0.156 \nl 
A3562 & --22.579 & 0.70 & --0.26 & 0.07 & 0.120 & 0.179 \nl 
A3565 & --22.619 & 0.53 & --0.21 & --2.14 & --0.048 & 0.164 \nl 
A3571 & --22.900 & 1.10 & --0.89 & 0.44 & 0.081 & 0.422 \nl 
A3574 & --22.539 & 0.74 & --0.09 & --1.58 & --0.113 & 0.172 \nl 
A3581 & --22.219 & 0.61 & --0.30 & --0.64 & 0.429 & 0.159 \nl 
A3716 & --22.567 & 0.71 & --0.34 & --0.35 & 0.085 & 0.169 \nl 
A3733 & --22.163 & 0.64 & 0.00 & --0.81 & 0.485 & 0.160 \nl 
A4038 & --22.342 & 0.45 & --0.38 & --0.11 & 0.118 & 0.174 \nl 
A4059 & --22.908 & 0.89 & --0.47 & 0.07 & --0.254 & 0.273 \nl 
\enddata 
\tablecomments{The BCG is assumed to lie at a distance given by its 
velocity in the Local Group frame with $H_0 = 80$ km~s$^{-1}$ 
Mpc$^{-1}$. $L_{44}$ is the X--ray luminosity in units of $10^{44}$ 
ergs s$^{-1}$ in the 0.1 -- 2.4 keV band. } 
\end{deluxetable} 

\end{document}